\font\ro=cmsy10                          % font with rope
\def\kcr{{\hbox{\ro \char'170}}}                % right-handed rope
\def\ktl{{\hbox{\ro \char'170}}}        % top end for left-handed rope
\def\ktr{{\hbox{\ro \char'170}}}        % " right
\def\kbl{{\hbox{\ro \char'170}}}        % " bottom left
\def\kbr{{\hbox{\ro \char'170}}}        % " right
\documentclass[12pt]{article}

\usepackage{a4}
\usepackage{amsfonts,amsmath,amssymb}
\usepackage{color}
\usepackage[dvips]{graphicx}
\usepackage{dcolumn,epsfig}
\usepackage{hyperref}

\makeatother

\def\un#1{\relax\ifmmode\@@underline#1\else
        $\@@underline{\hbox{#1}}$\relax\fi}

\let\du=\du                     % dot-under
\def\a{\alpha}
\def\b{\beta}

\def\d{\delta}

\def\f{\phi}
\def\g{\gamma}
\def\h{\eta}

\def\j{\psi}
\def\k{\kappa}
\def\l{\lambda}
\def\m{\mu}
\def\n{\nu}
\def\o{\omega}
\def\p{\pi}

\def\r{\rho}
\def\s{\sigma}

\def\F{\Phi}

\def\O{\Omega}

\def\ve{\varepsilon}

\def\ce{{\cal E}}

\def\cg{{\cal G}}

\def\car{{\cal R}}

\def\cw{{\cal W}}

\def\cz{{\cal Z}}

\def\bo{{\raise-.3ex\hbox{\large$\Box$}}}               % D'Alembertian
\def\pa{\partial}                                       % curly d
\def\de{\nabla}                                         % del
\def\TH{{\raise.2ex\hbox{$\displaystyle \bigodot$}\mskip-4.7mu \llap H \;}}
\def\face{{\raise.2ex\hbox{$\displaystyle \bigodot$}\mskip-2.2mu \llap {$\ddot
        \smile$}}}                                      % happy face
\def\leftrightarrowfill{$\mathsurround=0pt \mathord\leftarrow \mkern-6mu
        \cleaders\hbox{$\mkern-2mu \mathord- \mkern-2mu$}\hfill
        \mkern-6mu \mathord\rightarrow$}
\def\dvec#1{\vbox{\ialign{##\crcr
        \leftrightarrowfill\crcr\noalign{\kern-1pt\nointerlineskip}
        $\hfil\displaystyle{#1}\hfil$\crcr}}}           % <--> accent
\def\dt#1{{\buildrel {\hbox{\LARGE .}} \over {#1}}}     % dot-over for sp/sb
\def\sfrac#1#2{{\vphantom1\smash{\lower.5ex\hbox{\small$#1$}}\over
        \vphantom1\smash{\raise.4ex\hbox{\small$#2$}}}} % alternate fraction
\def\bfrac#1#2{{\vphantom1\smash{\lower.5ex\hbox{$#1$}}\over
        \vphantom1\smash{\raise.3ex\hbox{$#2$}}}}       % "
\def\afrac#1#2{{\vphantom1\smash{\lower.5ex\hbox{$#1$}}\over#2}}    % "
\def\on#1#2{\mathop{\null#2}\limits^{#1}}               % arbitrary accent
\def\[{\lfloor{\hskip 0.35pt}\!\!\!\lceil}
\def\]{\rfloor{\hskip 0.35pt}\!\!\!\rceil}
\def\Lag{{\cal L}}
\def\du#1#2{_{#1}{}^{#2}}
\def\ud#1#2{^{#1}{}_{#2}}

\def\fracm#1#2{\hbox{\large{${\frac{{#1}}{{#2}}}$}}}
\def\ha{{\fracmm12}}
\def\tr{{\rm tr}}

\def\un{\underline}
\def\fracmm#1#2{{{#1}\over{#2}}}

\def\low#1{{\raise -3pt\hbox{${\hskip 0.75pt}\!_{#1}$}}}

\def\Dot#1{\buildrel{_{_{\hskip 0.01in}\bullet}}\over{#1}}
\def\dt#1{\Dot{#1}}

\newskip\humongous \humongous=0pt plus 1000pt minus 1000pt
\def\caja{\mathsurround=0pt}
\def\eqalign#1{\,\vcenter{\openup2\jot \caja
        \ialign{\strut \hfil$\displaystyle{##}$&$
        \displaystyle{{}##}$\hfil\crcr#1\crcr}}\,}
\newif\ifdtup

\newcommand{\be}{\begin{equation}}
\newcommand{\ee}{\end{equation}}
\newcommand{\ba}{\begin{eqnarray}}
\newcommand{\ea}{\end{eqnarray}}
\newcommand{\nbe}{\begin{equation*}}
\newcommand{\nee}{\end{equation*}}

\newcommand{\lb}{\label}

\def\border{                                            % border
        \setlength{\unitlength}{1mm}
        \newcount\xco
        \newcount\yco
        \xco=-21
        \yco=19
        \begin{picture}(140,0)
        \put(\xco,\yco){$\ktl$}
        \advance\yco by-1
        {\loop
        \put(\xco,\yco){$\kcr$}
        \advance\yco by-2
        \ifnum\yco>-220
        \repeat
        \put(\xco,\yco){$\kbl$}}
        \xco=158
        \yco=19
        \put(\xco,\yco){$\ktr$}
        \advance\yco by-1
        {\loop
        \put(\xco,\yco){$\kcr$}
        \advance\yco by-2
        \ifnum\yco>-220
        \repeat
        \put(\xco,\yco){$\kbr$}}
        \put(-20,20){\tiny **University of Maryland * Center for String and
         Particle  Theory* Physics Department***University of Maryland *Center
        for String and Particle  Theory** }
        \put(-20,-221.5){\tiny **University of Maryland * Center for String and
         Particle  Theory* Physics Department***University of Maryland *Center
        for String and Particle  Theory** }
        \end{picture}
        \par\vskip-8mm}
\def\bordero{                                           % alternate border
        \setlength{\unitlength}{1mm}
        \newcount\xco
        \newcount\yco
        \xco=-31
        \yco=04
        \begin{picture}(140,0)
        \put(\xco,\yco){$\ktl$}
        \advance\yco by-1
        {\loop
        \put(\xco,\yco){$\kclr}
        \advance\yco by-2
        \ifnum\yco>-240
        \repeat
        \put(\xco,\yco){$\kbl$}}
        \xco=151
        \yco=04
        \put(\xco,\yco){$\ktr$}
        \advance\yco by-1
        {\loop
        \put(\xco,\yco){$\kcr$}
        \advance\yco by-2
        \ifnum\yco>-240
        \repeat
        \put(\xco,\yco){$\kbr$}}
\put(-20,12){\ooo bacdefghidfghghdhededbihdgdfdfhhdheidhdhebaaahjhhdahba

hgdedge
   hgfdiehhgdigicba}
\put(-20,-241.5){\ooo ababaighefdbfghgeahgdfgafagihdidihiidhiagfedhadbfd

ecdcdfa
   gdcbhaddhbgfchbgfdacfediacbabab}
        \end{picture}
        \par\vskip-8mm}
\def\headpic{                                           % UM heading
        \indent
        \setlength{\unitlength}{.4mm}
        \thinlines
        \par
        \begin{picture}(29,16)
        \put(165,16){\line(1,0){4}}
        \put(170,16){\line(1,0){4}}
        \put(180,16){\line(1,0){4}}
        \put(175,0){\line(1,0){4}}
        \put(180,0){\line(1,0){4}}
        \put(185,0){\line(1,0){4}}
        \put(169,0){\line(0,1){16}}
        \put(170,0){\line(0,1){16}}
        \put(179,0){\line(0,1){16}}
        \put(180,0){\line(0,1){16}}
        \put(184,0){\line(0,1){16}}
        \put(185,0){\line(0,1){16}}
        \put(169,16){\oval(8,32)[bl]}
        \put(170,16){\oval(8,32)[br]}
        \put(179,0){\oval(8,32)[tl]}
        \put(185,0){\oval(8,32)[tr]}
        \end{picture}
        \par\vskip-6.5mm
        \thicklines}
\def\title#1#2#3#4{\border\headpic {\hbox to\hsize{#4 \hfill UMDEPP #3}}\par
        \begin{center} \vglue .5in {\large\bf #1}\\[.6in]
        {#2}\\[.1in] {\it Department of Physics and Astronomy}\\
        {\it University of Maryland, College Park, MD 20742}\\[1.5in]
        {\bf ABSTRACT}\\[.1in] \end{center} \begin{quotation}}  % title stuff
\def\Title#1#2#3#4#5#6#7{\border\headpic
        {\hbox to\hsize{#7 \hfill UMDEPP #6}}\par
        \begin{center} \vglue .4in {\large\bf #1}\\[.4in]
        {#2}\\[.1in] {\it Department of Physics and Astronomy}\\
        {\it University of Maryland, College Park, MD 20742}\\[.1in]
        {#3}\\[.1in] {\it {#4}}\\ {\it {#5}}\\[.5in] {\bf ABSTRACT}\\[.1in]
        \end{center} \begin{quotation}}                 % " for 2 authors
\def\endtitle{\end{quotation}\newpage}        

\begin{document}

\thispagestyle{empty}

\def\dt#1{\on{\hbox{\bf .}}{#1}}                % (big) dot over
\def\Dot#1{\dt{#1}}

\def\gfrac#1#2{\frac {\scriptstyle{#1}}
        {\mbox{\raisebox{-.6ex}{$\scriptstyle{#2}$}}}}
\def\gg{{\hbox{\sc g}}}
\border\headpic {\hbox to\hsize{25th June, 2009 \hfill
{UMDEPP 09-040}}}
\par
{$~$ \hfill
{hep-th/0906.4978}}
\par

\setlength{\oddsidemargin}{0.3in}
\setlength{\evensidemargin}{-0.3in}
\begin{center}
\vglue .10in
{\large\bf Seeking the Loop Quantum Gravity Barbero-Immirzi \\ Parameter and Field  
                   in 4D, $\cal N$ = 1 Supergravity\footnote
{Supported in part  by National Science Foundation (Grants
PHY-0354401 \& PHY-0745779) }\  }
\\[.25in]

S.\, James Gates, Jr., \footnote{gatess@wam.umd.edu}
Sergei V. Ketov\footnote{ketov@phys.metro-u.ac.jp}
and Nicol\'as Yunes\footnote{nyunes@princeton.edu}
\\[0.6in]

{\it Center for String and Particle Theory\\
Department of Physics, University of Maryland\\
College Park, MD 20742-4111, USA}\\[0.25in]

{\it Department of Physics, Tokyo Metropolitan University\\
Minami-ohsawa 1-1, Hachioji-shi \\
 Tokyo 192-0397, Japan}\\[0.25in]

{\it Department of Physics, Princeton University\\
Princeton, NJ 08544, USA}\\[1in]

%---------------------------------------------------------------------------------
% Abstract
%---------------------------------------------------------------------------------

{\bf ABSTRACT}\\[.01in]
\end{center}
\begin{quotation}
{
We embed the Loop Quantum Gravity Barbero-Immirzi parameter and field within an action describing 4D,
$\cal N$ = 1 supergravity and thus within a Low Energy Effective Action of Superstring/M-Theory.
We use the fully gauge-covariant  description of supergravity in (curved) superspace.  The gravitational
constant is replaced with the vacuum expectation value of a scalar field, which in local supersymmetry is
promoted to a complex, covariantly chiral scalar superfield.  The imaginary part of this superfield
couples to a supersymmetric Holst term.  The Holst term also serves as a starting point in the Loop
Quantum Gravity action.   This suggest the possibility of a relation between Loop Quantum Gravity and
supersymmetric string theory, where the Barbero-Immirzi parameter and field of the former play the role of the
supersymmetric axion in the latter.  Adding matter fermions in Loop Quantum Gravity may require the
extension of the Holst action through the Nieh-Yan topological invariant, while in pure, matter-free
supergravity their supersymmetric extensions are the same.  We show that, when the Barbero-Immirzi
parameter is promoted to a field in the context of 4D supergravity, it is equivalent to adding a
dynamical complex chiral (dilaton-axion) superfield with a non-trivial kinetic term (or K\"ahler potential),
coupled to supergravity. 
}

${~~~}$ \newline

\endtitle

%---------------------------------------------------------------------------------
\section{Introduction}

Currently there are mainly two mathematical constructs which are often 
presented in adversarial positions to one another for the role of providing 
a road to a completely consistent quantum theory of gravity.  One of 
these, possessing a larger community of supporters among theoretical 
physicists, is Superstring/M-Theory (SSMT) and the other, with a 
corresponding smaller community of supporters, is Loop Quantum Gravity 
(LQG).  Theoretical physics is not a democracy as Nature possess the 
final and only vote to serve as the arbiter on the value of any theoretical 
construct. From this standpoint, the number of supporters is irrelevant 
when deciding which model better describes Nature.  
It is not the purpose of this work to debate whether 
one or the other of these proposal has been most 
successfully argued.  Relevant to this work, there have
appeared both arguments against the combination of these 
approaches~\cite{Nicolai:2005mc,Nicolai:2006id}, as well as
attempts to carry out such a combination~\cite{Jacobson:1987cj,Husain:1996gj,Ling:1999gn} 
To our knowledge, however, there is no completely 
rigorous exclusion principle that forbids the ultimate
theory of quantum gravity to be the result of a
combination of these approaches (and perhaps others),
which we believe makes it worthwhile to investigate 
possible regions of overlap.

LQG searches for the unification of General Relativity and Quantum 
Mechanics by postulating that spacetime itself is discrete~\cite{
Ashtekar:2004eh,Thiemann:2001yy, Rovelli:2004tv}. This model is 
formulated in terms of so-called {\emph{connection variables}}, two 
versions of which currently exist: the Ashtekar connection (selfdual 
$SL(2,C)$)~\cite{Ashtekar:1991hf} and the Ashtekar-Barbero connection (real 
$SU(2)$)~\cite{Barbero:1994ap}. Both formalisms can be derived 
from the Holst action which modifies the Einstein-Hilbert Action 
through the addition of a surface term that consists of the product of 
a parameter $\gamma$ (known as the Barbero-Immirzi, or BI, parameter) 
times the fully contracted dual Riemann curvature tensor~\cite{
Holst:1995pc}.

The BI parameter is a constant related to the minimum eigenvalue of 
the discrete area and discrete volume operators~\cite{Immirzi:1996di} 
and whose value is determined by the entropy-area relation when 
studying black hole thermodynamics.  Regardless  of the numerical 
choice of this parameter,  it does not affect the classical field equations 
since it appears solely through the addition of a surface term to the 
Einstein-Hilbert Action. 

Recently, however, the possibility of promoting this parameter to a 
scalar field (`scalarization') was proposed~\cite{Taveras:2008yf}. 
Replacing the BI parameter by the BI-field has the important
consequence that the modification of the Einstein-Hilbert action
is {\it no} longer by the addition of a pure surface term.
Instead, the field equations are now modified at the classical level, since integration 
by parts of the Holst term leads to corrections to the field equations 
that depend on derivatives of the BI field. One can easily show that 
this corrections reduce to GR in the presence of a scalar field 
stress-energy tensor~\cite{Taveras:2008yf} upon field redefinition. 
Shortly after its introduction, the canonical formulation of the quantum 
theory with a BI field was studied~\cite{Calcagni:2009xz}, followed 
by  its relation to the Peccei-Quinn mechanism~\cite{Mercuri:2009zi} 
and a possible topological interpretation~\cite{Mercuri:2009vk}. 

In the presence of space-time torsion (e.g., as induced by fermions), 
however, the Holst extension must be further augmented by the 
addition of another term to avoid classical modifications to the field 
equations.  Such an augmentation consists of adding a torsion squared 
term to the action, which thus transforms the Holst term to the so-called 
Nieh-Yan invariant~\cite{Nieh:1981ww}. When the BI parameter is 
promoted to a scalar field in the Nieh-Yan invariant, one finds 
generically that the Pontryagin topological density arises naturally 
in LQG~\cite{Mercuri:2009zt}. The effective action of LQG with a BI 
field is then equivalent to Dynamical Chern-Simons Modified Gravity 
(DCSMG)~\cite{Jackiw:2003pm, Alexander:2008wi,review}, which 
interestingly has already been connected to heterotic string theory~\cite{
Alexander:2004xd}. 

Though the possibility of any direct link between SSMT and LQG 
may be deemed remote by some, we believe it is of interest to probe 
for any such interconnection possessing the property of supersymmetry.  
Supergravity appears to be the natural crossing because, on the 
one hand side, supergravity is the least ad hoc way to include fermions 
into any field theory with gravity, while, on the other hand, supergravity 
arises as the low-energy effective action of superstrings. In this paper, 
we shall attempt to make the possibility of interconnections between 
LQG, DCSMG, and SSMT more clear and explicit, in the context of 4D, 
$\cal N$ =1 (pure) supergravity. Our choice of the four-dimensional 
(4D), $\cal N$ = 1 supergravity theory is for the sake of simplicity.

Let's first recall the basic (standard) motivation for supergravity. 
Supergravity is the theory of local supersymmetry and the latter is 
a symmetry that converts bosons into fermions and vice-versa. 
Supersymmetry seems well motivated in particles physics for 
unification of bosons and fermions, as well as for the consistency 
of SSMT. From the viewpoint of quantum gravity, supersymmetry 
is the tool for its consistent coupling to matter as well as the 
improvement of its ultraviolet (UV)-behavior (e.g., UV-divergences).   
Supergravity is the only known consistent route to couple spin-$3/2$ 
particles (gravitinos) to gravity in field theory.  In addition, supergravity 
(with matter) naturally arises as the low-energy effective action of 
SSMT. Such considerations therefore suggest that supergravity 
might also be unavoidable in LQG, if one desires to couple LQG to 
matter, and fermions in particular.   As regards discretizing 
supersymmetry on a lattice, see e.g., \cite{Catterall:2008dv,
Giedt:2008xm,Feo:2002yi}. 

The above relations are schematically portrayed in Fig.~\ref{diagram}. 
LQG and modified Holst or modified Nieh-Yan gravity are connected 
through the promotion of the BI parameter into a scalar field.   In turn, 
we here show that modified Holst gravity also generically arises in 
SSMT, where the BI field is related to the string theory axion (for real
values of the BI parameter). In the presence of fermions, modified 
Nieh-Yan gravity can also be mapped to DCSMG. In turn, the latter 
has already been shown to be generically unavoidable in string 
theory~\cite{Alexander:2004xd}.  In this way, we reinforce the possibility 
of  interconnection between the seemingly disparate LQG and SSMT.  
\begin{figure}
\begin{center}
\includegraphics[width=8cm,clip=true]{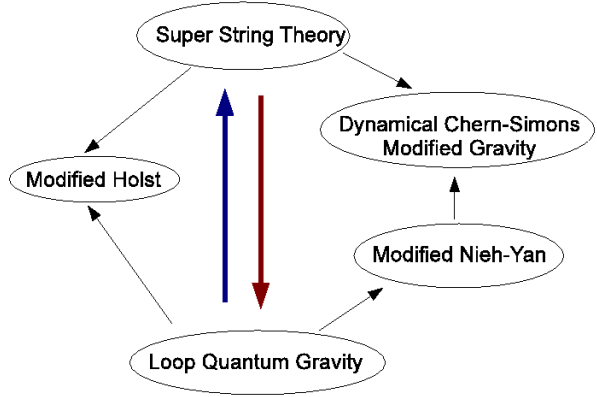}
\caption{\label{diagram} Schematic diagram of the relations drawn
in this paper between SSMT and LQG.}
\end{center}
\end{figure}

A deep comparison between LQG and SSMT requires not only 
the embedding of Holst or Holst-like actions in supergravity as discussed
in this paper, but also a mapping of the connection-triad variables
at the classical level and the background independent, non-Fock quantization
procedure at the quantum level to some SSMT equivalent. 
In all irreducible off-shell supergravity theories known, the connection variables 
do not appear in any superspace formulation. Thus, any attempt to 
reconcile LQG and SG theories begins with an intrinsic disagreement over the fundamental
variables of the theory. Relinquishing any of the requirements of irreducibility,   
off-shell structure or a superspace formulation, however, will open the way to the introduction  
of connection variables. Such issues are relegated to future work.

The remainder of this paper is divided as follows: 
Sec.~\ref{lqg} discusses the Holst action in LQG and its coupling of 
fermions with the Nieh-Yan invariant;
Sec.~\ref{supergravity-theory} describes the basics of 4D, $\cal N$ = 1 
supergravity in superspace;
Sec.~\ref{Holst:1995pc-par-in-supergravity} presents the relation 
between the BI parameter and 4D, $\cal N$ =1 supergravity;
Sec.~\ref{Holst:1995pc-field-in supergravity} shows how the BI field 
is related to superstring theory;
Sec.~\ref{nieh-yan-in supergravity} discusses the Nieh-Yan invariant 
in supergravity; 
Sec.~\ref{conclusions} concludes and points to future research.

We shall employ the following conventions in this paper: lower case 
{\it middle} greek letters $\m,\n,\ldots=0,1,2,3$ are used for curved 
spacetime vector indices; the lower case {\it middle} latin letters 
$m,n,\ldots=0,1,2,3$ are used for flat (target) space vector indices; the 
lower case {\it early} greek letters $\a,\b,\ldots=1,2$ are used for chiral 
spinor indices of one chirality, wherere the lower case {\it early} greek 
letters {\it with dots}, $\dt{\a},\dt{\b},\ldots=\dt{1},\dt{2}$ are used for 
chiral  spinor indices of the opposite chirality.  Symmetrization and 
antisymmetrization on the indices is denoted via $A_{(ab)} := ( A_{ab} 
+ A_{ba} )/2$ and $A_{[ab]} := ( A_{ab} - A_{ba} )/2$ respectively. We 
also use natural units where $\hbar=c=1$, and the gravitational 
coupling constant is introduced via $\k^2=8\p G_{\rm N}$.

%---------------------------------------------------------------------------------
\section{Loop Quantum Gravity}
\label{lqg}

%-------------------------------------
\subsection{The Holst Action}

The Holst action with the BI field is defined as the sum of three terms;
the Einstein-Hilbert term, a term involving the fully contracted dual 
Riemann tensor and terms involving coupling to matter.  Its explicit
form can be written as follows:
\ba
\label{Holst}
S_{\textrm{Holst}} &=& \frac{1}{4 \kappa^{2}} \int \epsilon_{mnpq} e^{m}
 \wedge e^{n} \wedge F^{pq}
\nonumber \\
&+& \frac{1}{2 \kappa^{2}} \int  \bar{\gamma} \; e^{m} \wedge e^{n}
 \wedge  F_{mn} + S_{\textrm{mat}},
\ea
where $\epsilon_{mnpq}$ is the Levi-Civita tensor, $e$ is the determinant 
of the tetrad (vierbein) $e^{m}$ and $e_{m}$ is its inverse, with spacetime 
indices here suppressed,  $F^{pq}$ is the curvature tensor constructed 
with the Lorentz spin connection $w^{mn}$, while $\bar{\gamma} = 1/\gamma$ 
is the inverse of the BI-field. We shall use exterior calculus in 
this section, and we refer the reader to~\cite{Taveras:2008yf,Romano:1991up} 
for details of its implementation in LQG. The quantity $S_{\textrm{mat}}$ 
represents additional matter degrees of freedom.

The field equations can be derived by varying the action of Eq.~\eqref{Holst} 
with respect to the tetrad and the connection. Variation with respect to the 
tetrad leads to the torsion condition
\be
2 T_{[m} \wedge e_{n]}  =  \frac{\partial_{q} \bar{\gamma}}{2 \bar{
\gamma}^{2} + 2} \left[ \epsilon_{rpmn} e^{r} \wedge e^{p} \wedge e^{q} 
-2 \bar{\gamma} e_{m} \wedge e_{n} \wedge e^{q} \right],
\label{torsion-cond}
\ee
which is solved to obtain
\be
T^{m} = \frac{1}{2} \frac{1}{\bar{\gamma}^{2} + 1} \left[ \epsilon^{m}{}_{npq} 
 \partial^{q}\bar{\gamma} + \bar{\gamma} \; \delta^{m}_{[n} \partial_{p]} \bar{
 \gamma}  \right] e^{n} \wedge e^{p}
\label{torsion-sol}
\ee
Clearly, in the limit as the BI field becomes constant one recovers the standard 
torsion-free limit of GR. 

Variation of the action with respect to the connection leads to the modified field 
equations. The on-shell field equations can be shown to reduce to the Einstein 
equations in the presence of a scalar field
\be
G_{\mu \nu} = k^{2} \left[ \left(\partial_{\mu} \phi\right) 
\left(\partial_{\nu} \phi\right) - \frac{1}{2}  
g_{\mu \nu} \left(\partial^{\sigma} \phi\right) \left(\partial_{\sigma} \phi\right) \right]
\label{holst-FE}
\ee
upon the field redefinition
\be
\phi = \sqrt{3} \sinh^{-1} \bar{\gamma}~.
\ee
Variation of the action with respect to this new field leads simply to the massless 
Klein-Gordon equation $\square \phi = 0$. When the BI field is a constant, the 
field equations classically reduce to the Einstein equations. 

The scalarization of the BI parameter has also been studied in the quantum 
theory~\cite{Calcagni:2009xz}. In fact, at a quantum level, the closing of the algebra requires that
the Ashtekar-Barbero connection be defined with a constant BI parameter~\cite{Calcagni:2009xz}. 
This constant value can be thought of as the expectation value of the BI field on the states
of the theory or as the asymptotic value of this field~\cite{simone-private} . 

The introduction of torsion to the connection suggests that higher-order curvature 
corrections to the action will lead to non-canonical kinetic terms for the scalar field. 
If the scalar field $\phi$ is then identified with the inflaton, one could possibly arrive 
at a realization of K-inflation from LQG~\cite{ArmendarizPicon:1999rj,Taveras:2008yf}. 
Of course, these comments are speculative as LQG does not require as of now 
higher-order curvature corrections to the Holst action.  Moreover, not just any 
modification to the kinetic energy of $\phi$ suffices to induce K-inflation, as 
precise conditions would have to be satisfied~\cite{Taveras:2008yf}.  

%-------------------------------------
\subsection{The Nieh-Yan Action}

The Holst action has been shown to lead to certain problems when 
coupling the theory to fermions.  Even when the BI parameter is treated as 
constant, fermion couplings have been shown to lead to torsion, which in turn 
leads to a classical effective action with fermion interaction terms that depend 
on the BI parameter at a classical level~\cite{Perez:2005pm,Freidel:2005sn}. 
Since the BI parameter is thought to be related to the quantum effect of spacetime
discretization, its appearance at  the level of an on-shell effective action has been thought somewhat 
problematic~\cite{Mercuri:2007ki,Mercuri:2006um}.   For this reason, it was 
suggested that the Holst term should by replaced by the Nieh-Yan invariant 
in the presence of fermions~\cite{Nieh:1981ww,Mercuri:2007ki,Mercuri:2006um}:
\ba
\label{NY}
S_{\textrm{Nieh-Yan}} &=& S_{\textrm{mat}} + 
 \frac{1}{4 k^{2}} \int \epsilon_{mnpq} e^{m} \wedge e^{n} \wedge F^{pq}
\\ \nonumber 
&+& \frac{1}{2 k^{2}} \int  \bar{\gamma} \; \left(e^{m} \wedge e^{n} \wedge  
F_{mn} - T^{m} \wedge T_{m} \right),
\ea
In the absence of fermions, this formulation is equivalent to the Holst one 
since torsion vanishes, thus also leading to the Ashtekar and Ashtekar-Barbero 
connections. In the presence of fermions, this formulation also leads to 
torsion, but the fermion interactions induced in the effective action are 
independent of the BI parameter~\cite{Mercuri:2007ki,Mercuri:2006um}.

Inspired by the scalarization of the BI parameter in the Holst action~\cite{
Taveras:2008yf}, there has been recently an effort to also study the 
scalarization of this parameter in the Nieh-Yan action.  When doing so, 
it was found again that variation of the action with respect to the tetrad 
leads to a torsion condition similar to Eq.~\eqref{torsion-cond}, whose 
solution is again non-vanishing torsion~\cite{Mercuri:2009zi}:
\be
T^{m}  = - \frac{1}{2} \epsilon^{m}{}_{npq} \left(\partial^{n} \bar\gamma\right) e^{p} \wedge e^{q} + \ldots
\ee
where the dots stand for possible fermion contributions. The inclusion of 
the $T^{m} \wedge T_{m}$ term in the action simplifies the torsion tensor, 
removing the prefactor in Eq.~\eqref{torsion-sol}.  Naturally then, the on-shell 
variation of the action with respect to the connection leads to the field 
equations of Eq.~\eqref{holst-FE} with the identification of $\phi \to \bar
\gamma$. Variation of the action with respect to the BI field reveals that
the BI field in the Nieh-Yan formulation satisfies a source-free, massless 
Klein-Gordon equation. 
 
One of the advantages of the Nieh-Yan action is the ability to consistently 
include fermions, which when combined with a BI field lead to BI field-fermion 
interactions that are proportional to $\star J_{A} \wedge d \bar\gamma$, 
where $J_{A}$ is the axial fermion current.   Upon integration by parts, 
one can relate such terms to $\bar\gamma \; R \wedge R$, where $R$ is 
the (torsion-free) Riemann tensor, because the axial fermion current is 
anomalous due to quantum effects~\cite{Mercuri:2009zi}.

The anomalous Pontryagin density ($R \wedge R$) can be used to 
cancel other CP violating terms that arise in gauge theories that are 
also proportional to this density. For example, 
such terms arise by the local gauge group due to the diagonalization 
of the quark mass matrices by a chiral rotation. Through the analog
of the Peccei-Quinn mechanism of quantum 
chromodynamics (QCD)~\cite{Peccei:1977hh,Mercuri:2009zi}, 
the BI field could be used to remove such CP violation from the action. 
In fact, the analogy to the Peccei-Quinn mechanism is so strong that
the BI field could be interpreted as the QCD axion to solve the 
strong CP problem through the chiral anomaly~\cite{Peccei:1977hh,Mercuri:2009zi}. 

The interaction terms, arising due to the coupling of fermions with the 
BI field, also naturally lead to a topological interpretation. Indeed, in 
Yang-MIlls theories, the requirement that states be invariant under 
large gauge transformations, leads to CP-violating anomalous terms 
of the form $\theta \; R \wedge R$, where $\theta$ is the Yang-Mills 
angle.  Such a term has been shown to also arise in LQG~\cite{
Ashtekar:1988sw}.  Since the BI field-fermion coupling leads to the 
same type of terms in the effective action, one can then identify the 
expectation value of the BI field with the $\theta$ angle~\cite{
Mercuri:2009vk}.  

The casting of the effective action of modified Nieh-Yan theory with 
a BI field in terms of the Pontryagin density leads to a interesting 
connection with DCSMG~\cite{Mercuri:2009zt}. Indeed, the latter 
is precisely defined by the sum of the Einstein-Hilbert action and 
the Pontryagin density multiplied by a dynamical field~\cite{
Jackiw:2003pm,Alexander:2008wi,review}. This connection becomes 
yet more interesting when one realizes that DCSMG contains terms 
that generically arise in the low energy limit of heterotic and type IIb 
string theories. In the latter, these terms arise due to a ten-dimensional 
generalization of the ABJ anomaly, required for the Ward identities to 
be preserved. This then suggests a deep connection between String 
Theory and LQG, which is at the heart of this paper, but before such 
a connection can be established,  we shall briefly review supergravity 
in superspace.

%---------------------------------------------------------------------------------
\section{4D, $\cal N$ = 1 supergravity in superspace}
\label{supergravity-theory}

A concise, manifestly supersymmetric and gauge-invariant description of 
supergravity is provided by {\it superspace} \cite{Gates:1983nr,
Wess:1992cp,Buchbinder:1995uq}.  The 4D, $\cal N$ =1 superspace is 
an extension of spacetime, parametrized by (bosonic) spacetime
coordinates $x^m$ and by extra fermionic (Grassmann) anti-commuting 
spinor coordinates $\theta_{\a}$ and $\bar{\theta}_{\dt{\a}}$.  General 
coordinate transformations of GR are extended in curved superspace to 
general supercoordinate transformations mixing $x$'s and $\theta$'s. 
A superfield amounts to a finite set (supermultiplet) of the ordinary fields 
(called superfield components) appearing as the coefficients in the 
superfield expansion in powers of $\theta$'s.  Superspace supergravity 
is the most natural way of unifying gravity and supersymmetry.  We refer 
the reader to the available textbooks \cite{Gates:1983nr,Wess:1992cp,
Buchbinder:1995uq,VanNieuwenhuizen:1981ae} for details about 
supergravity, superfields and their field components. In this section we 
formulate some basic ideas and give a few equations needed for the 
purposes of this paper.

An off-shell solution to the superspace Bianchi identities and the 
constraints, defining the $\cal N$ = 1 Poincar\'e-type minimal 
supergravity, gives rise to only three relevant tensor superfields, 
$\car$, $\cg_m$ and $\cw_{\a\b\g}$ (as parts of the supertorsion 
field). These fields are subject to the relations \cite{Gates:1983nr,
Wess:1992cp,Buchbinder:1995uq}
\ba 
\lb{Barbero:1994ap,Immirzi:1997di1}
 \cg_m &=& \bar{\cg}_m~,
 \nonumber \\
 \cw_{\a\b\g} &=& \cw_{(\a\b\g)}~,
 \nonumber \\
 \qquad
\bar{\de}_{\dt{\a}}\car &=&\bar{\de}_{\dt{\a}}\cw_{\a\b\g}=0~,
\ea
and
\ba 
\lb{Barbero:1994ap,Immirzi:1997di2}
 \bar{\de}^{\dt{\a}}\cg_{\a\dt{\a}} &=& \de_{\a}\car~,
 \nonumber \\
\de^{\g}\cw_{\a\b\g} &=&\frac{i}{2}\de\du{\a}{\dt{\a}}\cg_{\b\dt{\a}}+
\frac{i}{2}\de\du{\b}{\dt{\a}}\cg_{\a\dt{\a}}~~,
\ea
where $(\de\low{\a},\bar{\de}_{\dt{\a}}.\de_{\a\dt{\a}})$ represent the 
curved superspace $\cal N$ = 1  supercovariant derivatives, and bars 
denote complex conjugation.
For instance, the covariantly chiral complex scalar superfield $\car$ 
has the scalar curvature $R$ as the coefficient of its $\theta^2$ term. 
Similarly, the real vector superfield $\cg_{\a\dt{\a}}$ has the traceless 
Ricci tensor, $2 R_{(mn)} -\frac{1}{2}\h_{mn}R$, as the coefficient of 
its $\theta\s^m\bar{\theta}$ term\footnote{Here $\s^m=({\bf 1},i\vec{\s})$ 
stand for Pauli matrices.}.  The covariantly chiral, complex, totally 
symmetric, fermionic superfield $\cw_{\a\b\g}$ has the Weyl tensor 
$W_{\a\b\g\d}$ as the coefficient of its linear $\theta^{\d}$-dependent 
term. 

Gauge-fixed, off-shell, pure $\cal N$ = 1 supergravity can be also 
formulated in terms of the more conventional field components of 
supergravity superfields in spacetime (of Minkowski signature). We 
shall here consider the minimal (Poincar\'e) $\cal N$ = 1 supergravity 
in a {\it Wess-Zumino} (WZ) supersymmetric gauge \cite{Gates:1983nr,
Wess:1992cp,Buchbinder:1995uq}. The physical fields are a vierbein 
(or tetrad)  $e_{\m}^m(x)$ and a Majorana gravitino $\j_{\m}(x)$, while 
the auxiliary fields are a complex scalar $B(x)$ and a real vector $A_m
(x)$.\footnote{There are no physical degrees of freedom associated 
with auxiliary fields by definition, and thus, these fields are 
non-propagating. These fields are needed in supersymmetry, 
however, for an off-shell closure of the supersymmetry algebra and to 
make supersymmetry manifest. The auxiliary fields can become propagating 
in the presence of generic higher derivative modifications to the 
supergravity action.}  The tetrad is dimensionless, while the 
gravitino field is of canonical (mass) dimension $3/2$, while the auxiliary 
fields are of dimension $2$. All the auxiliary fields vanish on the pure 
supergravity equations of motion, ie. on-shell we have $B=A_m=0$.

The off-shell superspace constraints, defining a Poincar\'e supergravity, are necessary to reduce a curved 
superspace geometry to a geometry of supergravity. These constraints include the following: 
the {\it representation-preserving} constraints, which allow for the existence of covarianlty chiral superfields; 
the {\it conventional} constraints, which fix the vector covariant derivative in terms of the spinor ones and the 
spinor superconnections in terms of the spinor supervielbeins; and the {\it Einstein} (or WZ) constraints, which 
allow the passage from conformal supergeometry to Poincar\'e supergeometry in curved superspace. 
In particular, the minimal Poincar\'e supergravity constraints imply that 
the spin connection $\o_{\m}{}^{mn}(x)$ has the following structure 
\cite{Gates:1983nr,Wess:1992cp,Buchbinder:1995uq,VanNieuwenhuizen:1981ae}:
\be \lb{spinc}
\o_{\m mn} = \o_{\m mn}(e) +k_{\m mn}(\j) -\fracmm{2}{3}\ve_{\m mnp}A^p~~,
\ee
where $\o_{\m mn}(e)$ is the torsion-free spin connection of GR, 
and the gravitino-induced contorsion tensor $k_{\m mn}(\j)$ 
is given by
\be \lb{conto}
k_{\m mn} = \fracmm{\k^2}{4} \left( \bar{\j}_{m}\g_{\m}\j_n +
 \bar{\j}_{\m}\g_{m}\j_n - \bar{\j}_{\m}\g_{n}\j_m \right)~.
\ee 
In other words, we are in the 2nd-order formalism with a fixed spin connection  
$\o_{\m}{}^{mn}(e,\j,A)$, where we are now explicitly including spacetime indices. 
The physical (on-shell) spacetime torsion in supergravity is given by
\be \lb{tors}
 T_{\m\n}^p = \ha\nabla\low{[\m}(\o)e^p\low{\n]}
= \fracmm{\k^2}{4}\bar{\j}_{\m}\g^p\j_{\n}~, 
\ee
which obeys the identity
\be \lb{fierz}
\ve^{\m\n\l\r}T^p_{\m\n}T^p_{\l\r}=0~,\qquad {\rm or}\qquad \tr\,(T\wedge T)=0~, 
\ee
as the consequence of Fierz identities for the Majorana gravitino field. Hence, 
though the gravitino-induced torsion does not vanish in pure supergravity, its
contribution to the Nieh-Yan invariant in the last term of Eq.~({\ref{NY}) vanishes
on-shell because of Eq.~(\ref{fierz}).

The gauge-invariant action of pure supergravity in superspace is given by 
\cite{Gates:1983nr,Wess:1992cp,Buchbinder:1995uq}
\be \lb{psug} \eqalign{
S_{\rm sg}  &= -\fracmm{3}{\k^2}\int d^8z\, E^{-1} \cr
&= -\fracmm{3}{2\k^2} \int d^6z\, \ce \car + {\rm h.c.} }
\ee 
The form of the action on the first line of Eq.~(\ref{psug}) is the 
well-known, 
standard one of the superspace supergravity action \cite{Wess:1992cp} in 
terms of the supervielbein density  $E^{-1}={\rm sDet}\,E^M_A$ in {\it full} 
curved superspace\footnote{ In the next sections we are going to use the 
supergravity action in {\it chiral} curved superspace $z^{(4+2)}=(x,\theta)$ 
only, with the chiral supervielbein density $\ce$. The meaning of $x$ and 
$\theta$ is {\it different} in full and chiral superspace \cite{Gates:1983nr,
Wess:1992cp,Buchbinder:1995uq}.} $z^{(4+2+2)}=(x,\theta,\bar{\theta})$, 
where ${\rm sDet}$ is the superdeterminant.  The second line in 
Eq.~(\ref{psug}) is an equivalent representation of the same pure supergravity action 
in {\it chiral} superspace. The quantities ${\cal{E}}$ and ${\cal{R}}$ are 
supersymmetric generalizations of the volume element and the Lagrangian 
density in the chiral superspace.

The chiral superspace density (in WZ gauge) reads 
\be \lb{den}
\ce(x,\theta) = e(x) \left[ 1 -\k 2i\theta\s_m\bar{\j}^m(x) +
\k\theta^2 B(x)\right]~, \ee
where $e=\sqrt{-\det g_{\m\n}}$, $g_{\m\n}=\h_{mn}e^m_{\m}e^n_{\n}$ is a
spacetime metric of Minkowski signature, $\j^m_{\a}=e^m_{\m}\j^{\m}_{\a}$ 
is a chiral gravitino, and $B\equiv S-iP$ is the complex auxiliary field, with 
$S$ and $P$ defined as its real an imaginary parts respectively.

The chiral density integration formula, relating a curved superspace action
to the corresponding component action in spacetime, reads \cite{Gates:1983nr,
Wess:1992cp,Buchbinder:1995uq}
\be \lb{chiden}
  \int d^4x \; d^2\theta\,\ce \Lag =\int d^4x\, e\left\{ 
\Lag_{\rm last} +B\Lag_{\rm first}\right\}+ \ldots ~~,\ee 
where the dots stand for the gravitino-dependent terms. Here we have 
introduced the field components of the covariantly chiral superfield 
Lagrangian $\Lag(x,\theta)$ and $ \bar{\de}^{\dt{\a}}\Lag=0$  (the vertical 
bars denote the leading component of a superfield) as
\be \lb{com}
 \left. \Lag\right| =\Lag_{\rm first}(x)~,\qquad
 \left.\de^2\Lag\right|=\Lag_{\rm last}(x)~. \ee
Noting that ${\cal{L}} \propto {\cal{R}}$ in Eq.~(\ref{psug}), 
we must then compute ${\cal{R}}|$ and $\nabla^{2}{\cal{R}}|$, 
which reduce to~\cite{Gates:1983nr,Wess:1992cp,Buchbinder:1995uq}
\ba
\lb{part}
 \left.\car\right| &=& \frac{1}{3}\bar{B}=\frac{1}{3}(S+iP) 
 \\
 \lb{part2}
\left.\de^2\car\right| &=& \frac{1}{3}\left( R 
-\frac{i}{2}\ve\ud{mn}{pq}R\ud{pq}{mn}\right) +\frac{4}{9}\bar{B}B~,
\ea
where the spacetime curvature is given by 
\be \lb{curv}
R\ud{pq}{mn}= e_m^{\m}e_n^{\n} R\ud{pq}{\m\n} = e_m^{\m}e_n^{\n}
\left( \pa_{\m}\o_{\n}^{pq} +\o_{\m}^{pr}\o_{\n}^{rq} - 
\m \leftrightarrow\n \right)
\ee
and the scalar curvature is $R=e_m^{\m}e_n^{\n}R\ud{mn}{\m\n}~$. 

Combining all these ingredients, Eq.~(\ref{psug}) gives rise to the standard 
supergravity action in components in WZ gauge \cite{Gates:1983nr,
Wess:1992cp,Buchbinder:1995uq,VanNieuwenhuizen:1981ae}:
\ba
 \lb{comsg}
S_{\rm sg} &=& \int d^4x\,e\left\{ -\fracmm{1}{2\k^2} R\left[e,\o(e,\j)\right] 
\right. 
\\ \nonumber 
&+& \left.
\fracmm{i}{2}
\ve^{\m\n\l\r}\bar{\j}_{\m}\g_5\g_{\n}\nabla_{\l}\j_{\r} -\fracmm{1}{3}\bar{B}B
+\fracmm{4}{3}A^nA_n \right\}~.
\ea
The auxiliary terms $B\bar{B}$ and $A^{n} A_{n}$ arise here from the 
integration formula in Eq.~\eqref{part2}and from the decomposition of 
the spin connection in Eq.~\eqref{spinc} respectively.   The dual Riemann 
contribution in Eq.~\eqref{part2} cancels in the action of Eq.~(\ref{psug}) 
because of the identities for the torsion-free Riemann tensor.

%---------------------------------------------------------------------------------
\section{Holst action in supergravity with a BI parameter}
\label{Holst:1995pc-par-in-supergravity}

The gauge-invariant supergravity action in Eq.~(\ref{psug}) in chiral 
superspace can be slightly generalized by {\emph{complexifying}} 
the gravitational coupling constant in front of the first term via
\be \lb{cogr}
\fracmm{1}{\k^2} \to \fracmm{1}{\k^2}(1+ i\eta)
\ee
with the new dimensionless real parameter $\eta$. Such a complexification makes
 sense in supersymmetry since the second line of Eq.~(\ref{psug}) is the 
sum of a term (which is in general complex) and its complex conjugate. 

The complexification gives rise to a modified superfield supergravity action 
of the form
\be \lb{cogr2}
S_{\rm sgH}= -\fracmm{3}{2\k^2} \int d^6z\, \ce (1+i\h) \car + {\rm h.c.} 
\ee
which induces new terms proportional to $\eta$. When using the component
expansion of the WZ-gauge-fixed supergravity superfields (see e.g., Sec.~5.8
of Ref.~\cite{Buchbinder:1995uq}), one can find these {\it extra} 
$\h$-dependent terms 
\be
\lb{comsH}
S_{\rm sH} =  -\frac{\eta}{2} \int d^4x\,  \ve^{\m\n\l\r} \left[
\fracmm{1}{\k^2} R_{\l\r\m\n}(e,\o) 
+ \bar{\j}_{\m}\g_{\n}\nabla_{\l}\j_{\r}
\right] 
\ee
The $\eta$-independent terms in Eq.~(\ref{cogr2}) 
[not shown in Eq.~\eqref{comsH}] are just the standard supergravity action. 

The first gravitational term of Eq.~\eqref{comsH} coincides with the second 
term of the Holst action in the second line of Eq.~\eqref{Holst}~\cite{
Holst:1995pc} with a constant BI parameter. This identification allows us to 
relate the reciprocal of $\eta$ with $\gamma$, namely $\bar{\gamma} = 
\eta = \gamma^{-1}$~\cite{Barbero:1994ap,Immirzi:1997di}.  Therefore, the 
action
\be \lb{sH}
S_{\rm sgH}= -\fracmm{3i\eta}{2\k^2}\left( \int d^6z\, \ce \car - {\rm h.c.}
\right) 
\ee
is to be identified with the gauge-invariant and manifestly supersymmetric 
extension of the second term of Eq.~\eqref{Holst} in superspace. In WZ  
gauge, Eq.~(\ref{comsH}) just represents the relevant (i.e.~depending 
upon physical fields only) part of that supersymmetric extension in 
components, in agreement with the first calculation of the component 
supersymmetric Holst action (without auxiliary fields)~\cite{Tsuda:1999bg}. 
The auxiliary fields of the {\it pure} $\cal N$ = 1 supergravity theory vanish 
on-shell (in the absence of supersymmetric matter). The off-shell superfield 
action in Eq.~(\ref{cogr2}), which represents the full supersymmetric 
extension of the Holst action, depends, of course, on the auxiliary fields.

The original significance of the bosonic part of the action in 
Eq.~(\ref{comsH}) stems from dealing with the reality conditions required in the 
Ashtekar formulation of Euclidean quantum gravity \cite{Ashtekar:1991hf}. 
Since we work with a Minkowski signature, our BI parameter $\eta$ must 
be real, in accordance with the Ashtekar-Barbero formulation of LQG. In Euclidean 
space, however, $\eta$ is purely imaginary, and the (anti)-self-dual case, 
$\eta=\pm i$ just leads to the Ashtekar formulation of LQG.

The purely gravitational Lagrangian $\ve^{\m\n\l\r}R_{\l\r\m\n}(e,
\o(e))$ with the standard (no-torsion) connection $\o_{\m}{}^{mn}(e)$ 
vanishes because of Bianchi identities for the curvature. In supergravity, 
with the gravitino-induced torsion and $\o_{\m}{}^{mn}(e,\j)$ of 
Eq.~(\ref{spinc}), the supersymmetric Holst-Tsuda Lagrangian in Eq.~(\ref{comsH}) 
in components amounts to a total derivative \cite{Tsuda:1999bg},
\be \lb{sHtd}
-\ha\eta \ve^{\m\n\l\r}\pa_{\l}\left(\bar{\j}_{\m}\g_{\n}\j_{\r}\right)
= \fracmm{2\eta}{\k^2} \ve^{\m\n pq}\pa_{\m}T_{\n pq}~,\ee
where we have used the spacetime torsion tensor of Eq.~(\ref{tors}). 
Therefore, adding the supersymmetric (Holst) term with a constant BI 
parameter to pure $\cal N$ = 1 supergravity has no effect on its 
supergravitational equations of motion (though it is relevant in 
quantum supergravity), in agreement with Refs.~\cite{Tsuda:1999bg,
Kaul:2007gz}. 

The independence of the supergravitational equations of motion upon 
a constant BI parameter in the {\it extended} $\cal N$ = 2 and $\cal N$ = 
4 pure supergravities, without their coupling to supermatter (and for $\cal 
N$ = 1 as well) was  established by Kaul \cite{Kaul:2007gz}. Our analysis 
in $\cal N$ = 1 superspace confirms those conclusions in the case of pure 
$\cal N$ = 1 superegravity without its coupling to $\cal N$ = 1 matter 
superfields. In a matter-coupled $\cal N$ = 1 supergravity, the supergravity 
auxiliary fields couple to matter fields, so that  they do not vanish 
generically. Accordingly, the BI-dependence (or -independence) of the equations 
of motion in supergravity theories coupled to matter at the classical level 
should be checked separately and is beyond the scope of this paper.

%---------------------------------------------------------------------------------
\section{Holst action in supergravity with a BI field}
\label{Holst:1995pc-field-in supergravity}

The gravitational constant can be thought of as the vacuum expectation 
value (VEV) of a real scalar field, which leads to the well-known {\it 
Brans-Dicke} (BD) gravitational theories. In $\cal N$ = 1 supersymmetry,
the BD real scalar field must be promoted to a complex chiral scalar 
superfield $\cz(x,\theta)$, for the same reasons that the gravitational 
constant was complexified in the previous section. The leading field 
component of this complex chiral scalar superfield is the complex scalar
\be \lb{dax}
\left. \cz(x,\theta)\right| =\f(x) +ia(x).
\ee
The real part $\f(x)$ of the complex BD scalar couples to the scalar curvature 
and, therefore, it should be identified with a {\it dilaton}. The imaginary part
 $a(x)$ of the BD complex scalar field couples to the 2nd term in the Holst action 
(the dual scalar curvature) and, therefore, it should be identified with an 
{\it axion}. 

The independence of the Holst action on a (constant) BI parameter is crucial
for claiming an associated {\it Peccei-Quinn} (PQ) symmetry. Such a symmetry
of the Holst action with a BI field (cf. Ref.~\cite{Mercuri:2009zi})
\be \lb{pq}
a(x) \to a(x) +{\rm const.}
\ee
is often taken as the defining property of the axion field. From this perspective, 
it is clear that the BI field should be identified with the supergravity axion. 

More specifically, let's replace the coupling constants in the supergravity 
action of Eq.~(\ref{cogr2}), inside the superspace integration, by a complex 
covariantly chiral BI (or BD) superfield $\cz(x,\theta)$ as
\be \lb{repl}
-\fracmm{3}{2k^2}(1+i\eta) \to \cz(x,\theta)~~.
\ee
Such a replacement gives rise to the modified $\cal N$ = 1 supergravity action
\be \lb{lmult}
 S_{\cz} = \int d^6z\,\ce \cz\car + {\rm h.c.}
\ee
in chiral $\cal N$ = 1 superspace. The action in Eq.~(\ref{lmult}) appears to be 
the very special case of the modified supergravity actions introduced recently 
in Refs.~\cite{Gates:2009hu,Ketov:2009wc} --- see e.g., Eq.~(5.19) in Ref.~\cite{
Gates:2009hu} --- with the superpotential $V=0$. We shall thus employ the 
results of~\cite{Gates:2009hu,Ketov:2009wc} when $V=0$.

A super-Weyl transform of the superfield action in Eq.~(\ref{lmult}) can also be 
performed entirely in superspace, ie.~with manifest local $\cal N$ = 1 supersymmetry. 
In terms of field components, the superfield Weyl transform amounts to a Weyl 
transform, a chiral rotation and a (superconformal) $S$-supersymmetry 
transformation \cite{Howe:1978km}.  The chiral density superfield $\ce$ is a 
chiral compensator of the super-Weyl transformation
\be \lb{swt}
\ce \to e^{3\F} \ce~, \ee
whose parameter $\F$ is an arbitrary covariantly chiral superfield, $\bar{\de
}_{\dt{\a}}\F=0$. Under the transformation (\ref{swt}) the covariantly chiral 
superfield $\car$ transforms as 
\be \lb{rwlaw}
\car \to e^{-2\F}\left( \car - \fracm{1}{4}\bar{\nabla}^2\right)
e^{\bar{\F}}~~.
\ee
The super-Weyl chiral superfield parameter $\F$ can be traded for the chiral
Lagrange multiplier $\cz$ by using a simple holomorphic gauge condition
\be \lb{ch2} \cz=\F~~. \ee

With all these ingredients at hand, the super-Weyl transform of the action in 
Eq.~(\ref{lmult}) results in the classically equivalent action   
\be \lb{chimat2}
S_{\F} =  \int d^4x \; d^4\theta\, E^{-1} e^{\F+\bar{\F}}
\left[ \F +\bar{\F}  \right]
\ee
in full curved superspace of $\cal N$ = 1 supergravity.
Equation (\ref{chimat2}) has the standard form of the action of a chiral matter
superfield coupled to supergravity \cite{Gates:1983nr,Wess:1992cp,
Buchbinder:1995uq},
\be \lb{stand}
S[\F,\bar{\F}]= \int d^4x d^4\theta\, E^{-1} \O(\F,\bar{\F})~~,
\ee  
with the kinetic potential $\O(\F,\bar{\F})$, where we have defined
\be \lb{spots}
\O(\F,\bar{\F}) =  e^{\F+\bar{\F}}
\left[ \F +\bar{\F} \right]~.
\ee 
The K\"ahler potential $K(\F,\bar{\F})$ is given by \cite{Wess:1992cp,
Buchbinder:1995uq}
\be \lb{kaehler}
 K = -3\ln(-\fracmm{\O}{3})\quad {\rm or} \quad \O=-3e^{-K/3}~.
\ee

The proposed action in Eq.~(\ref{lmult}) with the chiral BI superfield 
$\cz$ is classically equivalent to the standard action of a chiral matter 
superfield $\F$ coupled to the minimal $\cal N$ = 1 supergravity and 
having a non-canonical kinetic term, ie.~a non-trivial K\"ahler potential.
The equivalent action in Eq.~(\ref{stand}) may be suitable as the starting 
point for generating K-inflation in the BI-field-modified supergravity along 
the lines of~\cite{Taveras:2008yf}.

%---------------------------------------------------------------------------------
\section{Nieh-Yan Action in Supergravity with a BI Field}
\label{nieh-yan-in supergravity}

In this section we comment on the embedding of the Nieh-Yan action 
with a BI field into supersymmetry. The Nieh-Yan invariant was defined 
by the integrand of the last term in Eq.~\eqref{NY}, without the BI field. 
Because of the identity in Eq.~(\ref{fierz}) it is clear that the 
supersymmetric extension of the Nieh-Yan density should be {\it the same} as 
the supersymmetric extension of the Holst action, i.e. both are given by 
Eq.~(\ref{sH}) in superspace, up to an overall normalization factor, as far as
the {\it pure} supergravity with a {\it constant} BI term is concerned. Moreover, 
because of Eq.~(\ref{sHtd}), the Nieh-Yan density (neglecting the Einstein-Hilbert 
term) can be rewritten as a divergence of the axial gravitino current
$J^{\m}=\ha\ve^{\m\n\l\r}\bar{\j}_{\n}\g_{\r}\j_{\l}$,
\be \lb{axgc}
S_{\rm Nieh-Yan}=  -\ha \int d^4x\, \nabla_{\m}J^{\m}~,
\ee
where $\nabla_{\m}$ is the covariant derivative. Equation~\eqref{axgc} 
is again in agreement with Ref.~\cite{Kaul:2007gz}. In superspace 
supergravity, it appears to be the consequence of the on-shell identity
\be \lb{onid}
i(\nabla^2\car - {\rm h.c.}) \propto \nabla_{\m}J^{\m}~~,
\ee
where the h.~c.~stands for the hermitian conjugate of the preceding term.

The invariance of an action in supersymmetry and supergravity is {\it 
usually} defined modulo a surface term, or up to a total derivative in 
the Lagrangian. A supersymmetric completion of a topological term 
in supersymmetry is often ambiguous (ie.~defined up to a surface term) 
since it does not change the bosonic value of a topological action. The 
superspace approach can, nevertheless, be effectively used in selecting 
the most natural (or minimal) consistent supersymmetric topological 
action that should be used in quantum gravity. When a constant BI 
parameter is promoted to a (space-time-dependent) BI {\it field}, there 
is no ambiguity in defining a (non-topological) supersymmetric action.

When supergravity is coupled to matter superfields, as is the case with
a BI superfield, the supergravity auxiliary fields do not vanish, while
their (algebraic) equations of motion determine them in terms of the 
physical (supergravity and matter) fields. In particular, the matter 
contribution to the spin connection enters via the vector auxiliary field
in Eq.~({\ref{spinc}), giving rise to matter-generated contributions
to spacetime torsion and rendering Eq.~({\ref{fierz}) no longer valid
contrary to the pure supergravity case. As a result, the supersymmetric
Holst action in components gets modified by matter-dependent terms.
We believe that the modification should be precisely in the form of
the Nieh-Yan-type extra term, though we did not verify it by an explicit 
calculation. The reason for this belief is the PQ symmetry [e.~g.~Eq.~(\ref{pq})]
of the superspace formulation of Holst supergravity with the BI superfield:
the action in Eq.~(\ref{chimat2}) is invariant with respect to the transformations 
in~Eq.(\ref{pq}) as the former merely depends on the sum $(\F +\bar\F )$.

We have thus established a direct mathematical relation between the 
Nieh-Yan action with a BI field and the $\cal N$ = 1 supergravity action 
in superspace. This mapping relates the BI field $\bar{\gamma}$ to the 
supergravity axion. In the non-supersymmetric four-dimensional theory, 
when the axion is coupled to the Holst term only, its kinetic energy becomes 
non-canonical, while with the Nieh-Yan term one recovers the standard canonical 
kinetic term.  In the supersymmetric case of a BI superfield coupled to 
supergravity, we get the genuine (supersymmetric) non-linear sigma-model 
representing the BI kinetic terms in the action of Eq.~(\ref{chimat2}).  

%---------------------------------------------------------------------------------
\section{Conclusions}
\label{conclusions}

We have shown that the possibility of migrating concepts
and constructs across the LQG/SSMT boundary can yield interesting
perspectives on issues that can arise.  While there is currently no known
requirement of the presence of supersymmetry in the LQG program,
embedding LQG considerations within supergravity theory may offer
hints on the possibility of a synthesis, daunting though this may seem.
If supersymmetry were ever to be found as a requirement in the LQG
program, then the stage would be set for a fascinating line of investigation
on the issue of Bekenstein-Hawking black hole entropy within the
two distinct approaches represented by LQG and SSMT.   In the
former all indications are the axion must have a non-vanishing
VEV while in the latter no such requirement arises.

A connection between the BI field and the axion also leads to two possible
and interesting interpretations: either the BI field and the axion field are one and the 
same; or the BI field is another massless degree of freedom that 
happens to possess similar couplings to curvature relative to the 
axion coupling.  In the former case, one can then attempt to constrain 
the BI field by studying the effect of its potential on cosmological 
observables. Work along these lines is currently underway~\cite{simone}.

Should the identification of the BI field with the axion hold true generically, 
one may recall the possibility of dark matter composed at least in part of axions. 
In such a case, the BI-field/particle might become a viable candidate for dark matter.  

%---------------------------------------------------------------------------------
\section*{Acknowledgments}

We are grateful to Abhay Asthekar, Simone Mercuri and Victor Taveras for useful suggestions 
and comments. We acknowledge support of the National Science Foundation Grants 
PHY-0354401 \& PHY-0745779. This research was also supported in part 
by the endowment of the John S.~Toll Professorship and the University of
Maryland Center for String \& Particle Theory. S.K. and N. Y. give additional acknowledgment 
for the hospitality of the University of Maryland and the Center for String and Particle Theory.

%---------------------------------------------------------------------------------
% BIBLIOGRAPHY

\newpage

\bibliographystyle{ieeetr}
\bibliography{references}

\end{document}